\documentclass[aps,prl,twocolumn]{revtex4}
\usepackage{psfig}
\def\duzomniejsze{<\kern-.7mm<}
\def\duzowieksze{>\kern-.7mm>}

\def\textbf#1{{\bf #1}}
\def\beq{\begin{equation}}
\def\eeq{\end{equation}}
\def\be{\begin{equation}}
\def\ee{\end{equation}}
\def\ben{\begin{eqnarray}}
\def\een{\end{eqnarray}}
\def\beqa{\begin{eqnarray}}
\def\eeqa{\end{eqnarray}}
\def\eea{\end{array}}
\def\bea{\begin{array}}
\newcommand{\bei}{\begin{itemize}}
\newcommand{\eei}{\end{itemize}}

\def\>{\rangle}
\def\<{\langle}

\def\dt#1{{{\kern -.0mm\rm d}}#1\,}
\def\ypodpis{\raise4mm\hbox{$\omega$}}
\def \IF {I_{f}}
\def \ID {I_{l}}
\def \IG {I_{g}}
\def \IFP {I_{f}(Q)}
\def \IDP {I_{l}(Q)}
\def \ED {E_D}
\def \EF {E_c}
\def \ER {E_r}

\def\DD{\Delta}
\def\DF{\Delta_f}
\def\DFP{\Delta_f(Q)}
\def\clocc{LOCC}

\def \s {\,\,\,\,}
\def\rab{\varrho_{AB}}

\def\r{\rho}

\begin{document}
\title{Are there phase transitions in information space?}
\author{Jonathan Oppenheim$^{(1)(2)}$, Micha\l{} Horodecki$^{(2)}$,
and Ryszard Horodecki$^{(2)}$}

\affiliation{$^{(1)}$
Racah Institute of Theoretical Physics, Hebrew University of Jerusalem, Givat Ram, Jerusalem 91904, Israel}
\affiliation{$^{(2)}$Institute of Theoretical Physics and Astrophysics,
University of Gda\'nsk, Poland}

\begin{abstract}
The interplay between two basic quantities -- quantum communication and 
information -- is investigated.
Quantum communication is an important resource for quantum states shared by 
two parties and is directly related to entanglement. 
Recently, the amount of {\it local information} that can be drawn from a state has
been shown to be closely related to the non-local properties of the state. 
Here we consider both formation and extraction processes, 
and analyze informational resources as a function of quantum 
communication.
The resulting diagrams in {\it information space} allow us to observe 
phase-like transitions when correlations become classical.
\end{abstract}
\maketitle

Quantum communication (QC) --
the sending of qubits between two parties -- is a primitive concept in quantum information theory.
Entanglement cannot be created without it, and conversely, entanglement
between two parties can be used to communicate quantum information through teleportation
\cite{teleportation}.
The amount of quantum communication needed to prepare a state, and the amount of quantum communication that
a state enables one to perform, are fundamental properties of states shared by two parties. This amount,
is identical to the entanglement cost $\EF$ \cite{cost} and entanglement of
distillation $\ED$ \cite{BBPS1996} respectively.
Perhaps surprisingly, these two quantities are different
\cite{Vidal-cost2002}.
There are even states which are "bound" in the sense that quantum
communication is needed to create it, but
nothing can be obtained  from it \cite{bound}.
Yet QC is a distinct notion from entanglement.
For example, one may need to use a large amount of QC while creating some
state of low $E_c$, in order to save some other resource.
In the present paper we will consider such a situation. The second primitive
resource of interest will be {\it information} which quantifies how pure
a state is. The motivation comes from (both classical or quantum)
thermodynamics:  it is known that
bits that are in a pure  state
can be used to extract physical work from a single heat bath \cite{Szilard},
and conversely work is required to reset mixed states to a pure form
\cite{Landauer,Bennett82}.

For distant parties, in order to use information to perform such tasks,
it must first be localized.
In \cite{OHHH2001} we considered how much information (i.e. pure states)
can be localized (or drawn) from a state shared between two parties.
Thus far, the amount of information needed to prepare a state has not been considered, a possible
exception being in
\cite{Bennett-nlwe} where it was noted that there was a thermodynamical
irreversibility between preparation and measurement
for ensembles of certain pure product states.
However, given the central role of quantum communication and information,
it would be of considerable importance
to understand the interplay between these two primitive resources.
In this Letter, we attempt to lay the foundation for this study by examining how much information
is needed to prepare a state and how much can be extracted from it as a function of quantum communication. For a given
state, this produces a curve in {\it information space}.
The shapes of the curve fall into a number of
distinctive categories which classify the state and only a small number of parameters are needed to characterize them.
The curves for pure states can be calculated exactly, and they are
represented by a one parameter family of lines of constant slope.
The diagrams exhibit features
reminiscent of thermodynamics,
and phase-like transitions
(cf. \cite{dorit-phase}) are observed.

An important quantity that emerges in this study is the {\it information surplus $\DF$}. It quantifies the additional
information that is needed to prepare a state when quantum communication resources are minimized.
$\DF$ tells us how much information is dissipated
during the formation of a state and is therefore closely related to the
irreversibility of state preparation
and therefore, to the difference between the entanglement of distillation
and entanglement cost.
When it is zero, there is no irreversibility in entanglement manipulations.
Examples of states with $\DF=0$ include pure states, and states with an
optimal decomposition \cite{BBPS1996} which is locally orthogonal.

Consider two parties in distant labs, who can perform local unitary operations, dephasing\cite{dephasing},
and classical communication. It turns out to be simpler to
substitute measurements
with dephasing operations, since we no longer need to keep track of the informational cost
of reseting the measuring device. (This cost was noted
by Landauer \cite{Landauer} and used by Bennett \cite{Bennett82} to
exorcize Maxwell's demon.)
The classical communication channel can also be thought of as a dephasing channel. Finally, we
allow Alice and Bob to add noise (states which are proportional to the identity matrix) since pure noise contains no
information.
Note that we are only interesting in accounting for resources that are "used up" during the preparation procedure.
For example, a pure state which is used and then reset to its original state at the end of the procedure,
does not cost anything.

Consider now  the information extraction process of \cite{OHHH2001}.
If the two parties have access to a quantum channel,
and share a state $\rab$, they can extract all the information from the state
\beq
I=n-S(\rab)
\label{eq:info}
\eeq
where $n$ is the number of qubits of the state, and $S(\rab)$ is
its Von Neumann entropy.
Put another way, the state can be compressed, leaving $I$ pure qubits.
However, if two parties do not have access to a quantum channel, and
can only perform local operations and communicate classically (LOCC), then in general,
they will be able to draw less local information from the state. In
\cite{OHHH2001} we defined
the notion of the {\it deficit} $\DD$ to quantify the information that can no longer be drawn under LOCC.
For pure states, it was proven that $\DD$ was equal to the amount
of distillable entanglement in the state.

Let us now turn to formation processes and define $\DFP$ as follows. Given an amount of quantum
communication $Q$,
the amount of information (pure states) needed to prepare
the state $\rab$
under LOCC is given by $\IFP$.
Clearly at least $\EF$ bits of quantum communication are necessary.
In general, $\IFP$ will be greater than
the information content $I$.
The surplus is then
\beq
\DFP=\IFP-I \s .
\eeq

The two end points are of particular interest I.e. $\DF\equiv\DF(\EF)$ where quantum communication is minimized,
and $\DF(\ER)=0$ where we use the quantum channel enough times that $\IF(\ER)=I$. Clearly
$
\EF\leq \ER \leq min\{S(\r_A),S(\r_B)\}
$
where $\r_A$ is obtained by tracing out on Bob's system. This rough bound is obtained by noting that at a minimum,
Alice or Bob can prepare the entire system locally, and then send it to the other party through the quantum channel
(after compressing it). We will obtain a tight bound later in this paper.

The general procedure for state preparation is that Alice uses a 
{\it classical instruction set} (ancilla
in a classical state) with probability
distribution matching that of the decomposition which is optimal for a given $Q$. Since the instruction set contains
classical information, it can be passed back and forth between Alice and Bob.
Additionally they need $n$ pure standard states. The pure states are then correlated with the ancilla, and then sent.
The ancilla need not be altered by this procedure, and can be reset and then reused and so at worse we have
$\IF\equiv \IF(\EF)\leq n$
and
$
\DF \leq S(\rab).
$
We will shortly describe how one can do better by extracting information from correlations
between the ancilla and the state.

\begin{figure}
\psfig{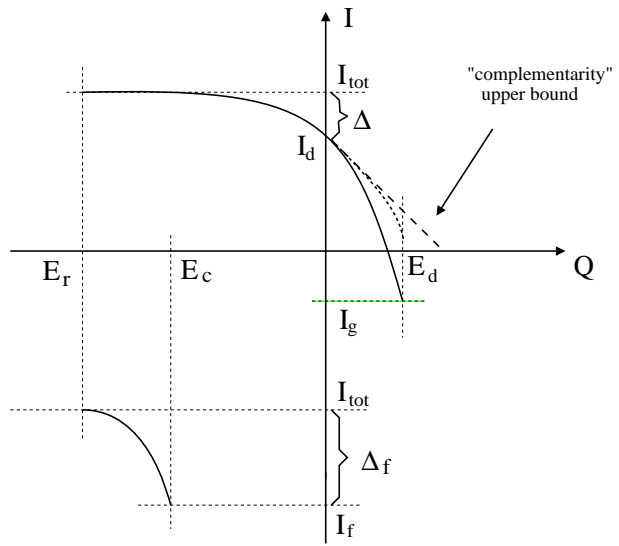}
\caption[Formation and extraction curves for a typical state]
{Formation and extraction curves for a  generic mixed state.
The short-dashed line represent the variant where
information can be extracted from the "garbage" left after
entanglement distillation ($I_g>0$).
In general, the curves need not be smooth.  The formation
curve is in the lower left quadrant.
\label{generic}}
\end{figure}

The pairs $(Q, \IFP)$ form curves in {\it information space}.
In Figure \ref{generic}
we show a typical curve which we now explain.
Since we will be comparing the formation curves to extraction curves, we will adopt the convention
that $\IFP$ and $Q$ will be plotted on the negative axis since we are using
up resources.
It can be shown that $\IFP$ is concave, monotonic and continuous.
To prove concavity, we take the limit of many copies of the state $\rab$. Then given
any two protocols, we can always toss a coin weighted with probabilities $p$ and $1-p$
and perform one of the protocols with this probability. There will always be a protocol which is at least
as good as this. Monotonicity is obvious (additional quantum communication can only help), and continuity
follows from monotonicity, and the existence of the probabilistic protocol.

The probabilistic protocol can be drawn as a straight line between the points $(\ER,\IF(\ER))$ and $(\EF,\IF(\EF))$.
There may however exist a protocol which has a lower $\IFP$ than this straight
line, i.e. the curve $\IFP$ satisfies
\beq
\IFP \leq I+\frac{I(\ER)-\IF(\EF)}{\EF-\ER}(Q-\ER)
\eeq

Let us now look at extraction processes. The idea is that we draw both local information (pure separable states), and
distill singlets. The singlets allow one to perform teleportations, so that we are in fact, extracting the potential
to use a quantum channel. We can also consider the case where we use the quantum channel to assist in the information extraction
process. We can therefore write the extractable information $\IDP$ as a function of $Q$. When $Q$
is positive, we distill singlets at the same time as drawing
information, and when $Q$ is negative, we are using the quantum
channel $Q$ times to assist in the extraction (see also Figure
\ref{generic}).

There appear to be at least three special points on the curve.
The first, is the point $\ID\equiv\ID(0)$.
This was considered in \cite{OHHH2001} when we draw maximal local information
without the aid of a quantum channel.
Another special point is the usual entanglement distillation procedure $\IG=\ID(\ED)$.
The quantity $\IG$ is the amount of local information
extractable from the "garbage" left over from distillation. $\IG$ can be negative as
information may need to be added to the system in order to distill all the
available entanglement.
Finally, $I=\ID(\ER)$ is the point where we use the quantum channel enough times
that we can extract all the available information. This is the same number of times that the
quantum channel is needed to prepare the state without any information surplus
since both procedures are now reversible.

\begin{figure}
\psfig{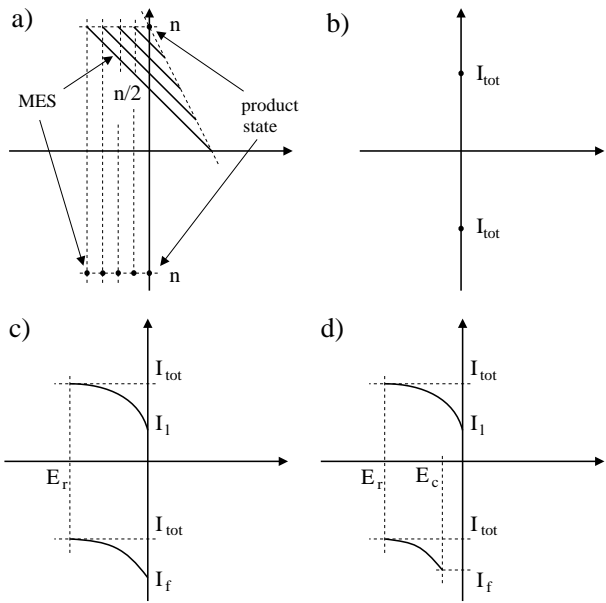}
\caption[l]{a) pure states b) states with $\DD=0$ c) separable
states with $\DD>0$ d) bound entangled states with $\DD>0$.
\label{various}
}
\end{figure}

Just as with the formation curve,
$\IDP$ is convex, continuous and monotonic.
For $Q\geq 0$ there is an upper bound on the extraction curve
due to the classical/quantum complementarity
of \cite{compl}.
\beq
I+Q \leq \ID
\label{eq:compl}
\eeq
It arises because one bit of local information can be drawn from each distilled
singlet, destroying the entanglement.
One might suppose that the complementarity relation (\ref{eq:compl})
can be extended into the region $Q< 0$. Perhaps surprisingly, this is not the case, and we have found
that a small amount of quantum communication can free up a large amount of information.
In Figure \ref{various}a we plot the region occupied by pure states.
For extraction processes, pure states
saturate the bound of Eq. (\ref{eq:compl}) \cite{compl}.
For formation processes they are represented as points.

In general, if $\DF=0$ then $\EF=\ED$. Examples include
mixtures of locally orthogonal pure states\cite{termo}.
The converse is not true, at least for single copies, as there
are separable states such as those of \cite{Bennett-nlwe} which have
$\DF\neq 0$, and
$\DD\neq 0$.

It therefore appears that $\DF$ is not a function of the entropy of the state, or of the entanglement,
but rather, shows how chaotic the quantum correlations are. It can also be thought of as the information
that is dissipated during a process, while $\DD$ can be thought of as the {\it bound information}
which cannot be extracted under LOCC.
Figure \ref{various}b-d shows the curves for some different types of states. It is
interesting the extent to which one can classify the different states just by examining the diagrams
in information space.

The quantities we are discussing have  (direct or metaphoric) connections with thermodynamics.
Local information can be used to draw physical work, and quantum communication has been
likened to quantum logical work \cite{termo}.
One is therefore tempted to investigate whether there can be some effects
similar to phase transitions.
Indeed, we will demonstrate such an effect for a family of mixed states
where the transition is of second order, in that the
derivative of our curves will behave in a discontinuous way.

To this end we need to know more about $\ER$ and $I_f$. Consider the
notion of {\it \clocc-orthogonality} (cf. \cite{termo}). One says that
$\varrho^i$ is \clocc-orthogonal, if by LOCC Alice and Bob
can transformed $\sum_ip_i|i\>_{A'}\<i|\varrho^i_{AB}$
into $|0\>_{A'}\<0|\otimes \sum_ip_i\varrho^i_{AB}$ and vice versa;
$|i\>_{A'}$ is the basis of Alice's ancilla. In other words, Alice
and Bob are able to correlate the state $\varrho_i$ to
orthogonal states of a local ancilla as well as reset the correlations.
Consider a state $\varrho_{AB}$ that can be \clocc-decomposed,
i.e. it is a mixture of \clocc-orthogonal states
$\varrho=\sum_ip_i\varrho_i$. The decomposition suggests
a scheme for reversible formation of $\varrho$. Alice prepares locally the
state $\varrho_{A'AB}=\sum_ip_i|i\>_{A'}\<i|\varrho^i_{AB}$. This costs
$n_{A'AB}-S(\varrho_{A'AB})$ bits of information. Conditioned on $i$,
Alice compresses the halves of $\varrho_i$, and sends them to Bob via a
quantum channel. This costs $\sum_ip_i S(\varrho_B)$ qubits of quantum
communication. Then, since the $\varrho_i$ are \clocc-orthogonal, Alice and
Bob can reset the ancilla, and return $n_{A'} $ bits. One then finds,
in this protocol, formation costs $n_{AB}-S(\varrho_{AB})$
bits, hence it is reversible. Consequently $\ER\leq \sum_ip_i S(\varrho_B)$,
hence
\be
\ER(\rab) \leq \inf \min_X\sum_ip_i S(\varrho_X^i)
,\quad X=A,B
\label{eq:er-bound}
\ee
where the infimum runs over all \clocc-orthogonal decompositions of $\rab$.

We can also estimate $\IF$ by observing
that the optimal decomposition for
entanglement cost is compatible with \clocc-orthogonal decompositions, i.e.
it is of the form $\{ p_i q_{ij}, \psi_{ij}\}$ where
$\sum_j q_{ij} |\psi_{ij}\>\<\psi_{ij}|=\varrho_i$. Now, Alice prepares
locally the state
$\varrho_{A'A''AB}=\sum_ip_iq_{ij} |i\>_{A'}\<i|\otimes
|j\>_{A''}\<j| \otimes |\psi_{ij}\>_{AB}\<\psi_{ij}|$. Conditional on
$ij$, Alice compresses the halves of $\psi_{ij}$'s and sends them to Bob.
This costs on average $\EF$ qubits of communication. So far Alice
borrowed $n_{A'A''AB}- S(\varrho_{A'A''AB})$ bits. Alice and Bob then
reset and return ancilla $A'$ (this is possible due to \clocc-orthogonality
of $\varrho_i$) and also return ancilla $A''$ without resetting.
The amount of bits used is $n_{AB} - (S(\varrho_{AB})-
\sum_ip_iS(\varrho_i))$, giving
\be
\DF\leq \inf\sum_ip_iS(\varrho_i)\leq S(\varrho)
\label{eq:Delta-bound}
\ee
where, again, the infimum runs over
the same set of decompositions as in Eq. (\ref{eq:er-bound}) providing
a connection between $\DF$ and $\ER$. In the procedure above,
collective operations were used only in the compression stage.
In such a regime the above bounds are tight.
There is a question, whether by some sophisticated collective
scheme, one can do better. We conjecture that it is not the case,
supported by the remarkable result of \cite{HaydenJW2002}.
The authors show that an ensemble of nonorthogonal states
cannot be compressed to less than $S(\varrho)$ qubits
even at the expense of classical communication. In our case
orthogonality is replaced by \clocc-orthogonality, and
classical communication by resetting. We thus assume equality in
Eqs. (\ref{eq:er-bound}), (\ref{eq:Delta-bound}). Thus
for a state that is not \clocc-decomposable (this holds for all two
qubit states that do not have a product eigenbasis) we
have $\DF=S(\rab)$, $\ER=\min\{S(\varrho_A),S(\varrho_B)\}$.

\begin{figure}
\psfig{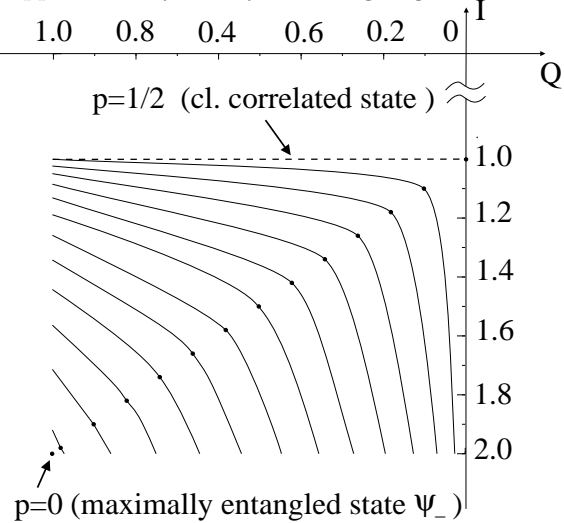}
\caption{``Phase transition'' in the family of states of Eq. (\ref{eq:bellmix})
\label{fig:phase} }
\end{figure}

Having fixed two extremal points of our curves, let us see
if there is a protocol which is better than the probabilistic one
(a straight line on the diagram).
We need to find some intermediate protocol which is cheap in
both resources. The protocol is suggested by the decomposition
$\varrho=\sum_ip_i\varrho_i$ where $\varrho_i$ are themselves
\clocc-orthogonal mixtures of {\it pure} states.
Thus Alice
can share with Bob each $\varrho_i$ at a communication cost of
$Q=\sum_ip_i \EF(\varrho_i)$. If the states $\varrho_i$
are not \clocc-orthogonal, Alice cannot reset the instruction set,
so that the information cost is $I=n-\sum_ip_iS(\varrho_i)$. We will now show
by example, that this may be a very cheap scenario. Consider
\be
\varrho= p|\psi_+\>\<\psi_+|+(1-p)|\psi_-\>\<\psi_-|,
\quad p\in[0,{1\over 2}]
\label{eq:bellmix}
\ee
with $\psi_\pm={1\over \sqrt2}(|00\>\pm|11\>)$.
When $p\not=0$ we have $\ER=1$, $\IF=2$,
$E_c=H({1\over 2} + \sqrt{p(1-p)})$ \cite{Vidal-cost2002} where
$H(x)=-x\log x - (1-x) \log (1-x)$ is the binary entropy; thus our extreme points
are $(1,2-H(p))$ and $(E_c, 2)$. For $p=0$ the state has $\DD=0$
hence the formation curve is just a point. We can however plot it as
a line $I=1$ (increasing $Q$ will not change $I$). Now,
we decompose the state as $\varrho=2p \varrho_s + (1-2p)|\psi_-\>\<\psi_-|$,
where $\varrho_s$ is an equal mixture of \clocc-orthogonal states $|00\>$ and $|11\>$.
The intermediate point is
then $(1-2p,2-2p)$. A family of diagrams with changing parameter $p$ is
plotted in Fig. (\ref{fig:phase}). The derivative
$\chi(Q)={\partial Q\over \partial I}$ has a singularity at $p=1/2$. Thus we
have something analogous to a second order phase transition.
The quantity $\chi(Q)$ may be analogous to a quantity such as the magnetic susceptibility.
The transition is between states having $\DD=0$ (classically correlated)\cite{OHHH2001} and states
with $\DD\not=0$ which contain quantum correlations. It would be
interesting to explore these transitions and diagrams further, and
also the trade-off between information and quantum communication. To
this end, the quantity $\DFP+Q-\EF$ appears to express this tradeoff.
Finally, we hope that the presented approach may clarify an intriguing notion in
quantum information theory,  known as the {\it thermodynamics of
entanglement} \cite{popescu-rohrlich,termo,thermo-ent2002}.

{\bf Acknowledgments}:
This work supported by EU grant EQUIP, Contract IST-1999-11053.
J.O. acknowledges
Lady Davis, and
grant 129/00-1 of the ISF. M.H. and R.H. thank C. Tsallis for discussions.

\begin{thebibliography}{17}
\bibitem[{dep()}]{dephasing}
\bibinfo{note}{Dephasing can be written as $\sum_i P_i \r_{AB} P_i$ where the
  $P_i$ are orthogonal local projection operators}.
\end{thebibliography}

\end{document}